\documentstyle[12pt]{article}

\begin{document}

\title{\bf KINK-LIKE SOLUTION FOR THE LORENTZ-VIOLATING $\phi^4$-THEORY EQUATION OF MOTION WITH DISSIPATION}

\author{M.A. Knyazev\\{\it Belarusian National Technical University,}\\{\it 65, Independence Ave., Minsk, 220013, Belarus}
\\{\it e-mail: maknyazev@bntu.by}}

\maketitle
\begin{abstract}

A (1+1)-dimension equation of motion for $\phi^4$-theory is
considered for the case of simultaneous taking into account the
processes of dissipation and violation the Lorentz-invariance. A
topological non-trivial solution of one-kink type for this equation
is constructed in an analytical form. The modified Hirota method for
a solving the nonlinear partial derivatives equations is applied. A
modification of the method led to the special conditions on
parameters of the model and solution.

{\bf Key words}: kink, Lorentz-violating model, $\phi^4$-theory,
dissipation, modified Hirota method.
\medskip
\\ PACS number(s): 02.30.Jr, 47.27.Ak, 52.25.-b

\end{abstract}

The Lorentz invariance of the field theory equations of motion is a
common place in modern physics. This takes a place for the
$\phi^4$-theory also. This theory is widely used in the different
fields of physics: classical and quantum field theories, physics of
the elementary particles, theory of the phase transitions in
condensed matter, physics of magnetism, inflation theory, theory of
formation the topological defects in cosmology and so on \cite {1}.

Last years the physical processes with a violation of the
Lorentz-invariance attract an attention of researches. One can
mention the Maxwell's electrodynamics which takes into account the
Chern-Simons interaction, an accounting the members violating
Lorentz invariance in a low-energetic limit of the standart model
and high-energetic limit in the string models, non-commutative field
theories, supersymmetric theories, a possible limits on the Lorentz
invariance from the radioastronomy and so on. In the scalar field
theory models, including multi-components models, an equation of
motion usually has a kink or kink-like solution. A violation of the
Lorentz invariance for such equation takes a place for the Kondo
effect, a capture of fermions, in the problems of entropy of various
informative processes and others. A review of the modern state of
art in this field are presented in paper \cite{2} (see also,
\cite{3}).

In this paper (1+1)-dimensional equations of motion in
$\phi^4$-theory are considered. Below for convenience where it is
possible the coefficients in all equations of motion will be equal
to unity.

The static equation of motion in $\phi^4$-theory has a form
\begin{equation} \label{EQ1}
\phi_{xx}+\phi-\phi^{3}=0,
\end{equation}
where $\phi_{xx}=\frac{\partial^2 \phi}{\partial x^2}$ and so on.
The solution of this equation is well-known \cite{4}
\begin{equation} \label{EQ2}
\phi (x)_{st} = \pm \tanh(\frac{x - x_{0}}{\sqrt{2}}).
\end{equation}
Here sign plus corresponds to kink and sign minus - to antikink.
$x_{0}$ is an initial phase and determines a coordinate of the
centrum of kink.

For dynamical case Eq.(1) takes a form
\begin{equation} \label{EQ3}
\phi_{tt}-\phi_{xx}-\phi+\phi^{3}=0.
\end{equation}
Eq.(3) is not an integrable one. There is only one-kink
(one-antikink) solution for this equation. There are no solutions
correspond to the coupled states of any kinks, antikinks or kinks
and antikinks \cite{5}, \cite{6}. Eq.(3) possesses the Lorentz
invariance. Hence, its solution may be written in a form
\begin{equation} \label{EQ4}
\phi (x,t)= \pm \tanh(\frac{x - x_{0} -ut}{\sqrt{2(1 - u^{2}}}),
\end{equation}
where $-1 < u < 1$ is a kink velocity.

If one takes into account a dissipation the equation of motion may
be written in a form
\begin{equation}  \label{EQ5}
\phi_{tt}-\phi_{xx}+\alpha \phi_{t}-\phi+\phi^{3}=0,
\end{equation}
where $\alpha$ is a damping coefficient. This equation is the
Lorentz invariant one. It is possible to solve Eq.(5) by using the
elliptic functions \cite{7} or the Hirota direct method for
nonlinear partial differential equations \cite{1}, \cite{6}. The
solution has a form
\begin{equation}  \label{EQ6}
\phi (x,t)= \frac{1}{2} \{1- \tanh[\alpha (x - ut)+x_{0}]\}.
\end{equation}
This solution is not a kink, but kink-like object. For Eq.(5) it is
also impossible to construct the solutions which correspond to
coupled states of any kinks, antikinks or kinks and antikinks.
Hence, Eq. (5) is also not integrable one.

Consider the equation of motion for $\phi^{4}$-theory in which the
Lorentz invariance is violated. This equation has a form
\begin{equation} \label{EQ7}
\phi_{tt}-\phi_{xx}+2 \beta\phi_{xt}-\phi+\phi^{3}=0.
\end{equation}
A constant coefficient $\beta > 0$ describes a violation of the
Lorentz invariance. A case $\beta = 0$ is trivial one. A coefficient
$2$ is introduced for a convenience of calculation. It is clear that
term $2 \beta\phi_{xt}$ does not play any role in a static case. In
paper \cite{3} it is shown that Eq.(3) is invariant under a
transformation
\begin{equation}  \label{EQ8}
x^{'}= \gamma (x_{0} - ut), \gamma = \frac{1}{\sqrt{1 - u^{2} + 2
\beta u}}.
\end{equation}
For such a case the topological nontrivial solution for Eq.(8) has a
form
\begin{equation} \label{EQ9}
\phi(x,t)=\tanh[\gamma (x - ut - x_{0})].
\end{equation}
A positive value of velocity correspond a motion of kink to the
right, a negative - to the left. The problem of integrability for
Eq.(9) is still open now.

Eqs.(3), (5) and (7) possesses a some similarity. At the same time
the differences between them consist in different linear terms.
Hence, these terms do not influence essentially on nonlinear
properties of equation and a form of their solutions. The solutions
of these equations are determined by a hyperbolic tangent and
represent a kink or kink-like object.

A further development of studies of the equation of movement for
$\phi^{4}$-theory consists in its generalization. Eqs. (1),(3), (5)
and (7) have the same potential. A modification of the potential by
taking into account a higher degrees of $\phi$ is one of the ways of
generalization. There are many papers on this topic (see, for
example, \cite{8}, \cite{9}). One another way of generalization is a
deformation of known potential \cite{10}, \cite{11}.

A generalization of the equation of motion in which a dissipation
and violation of the Lorentz invariance take into account
simultaneously have a significant interest also. Such equation may
be written in a form \cite{12}
\begin{equation} \label{EQ10}
\phi_{tt}-\phi_{xx}+\alpha \phi_{t}+2\beta
\phi_{xt}-\phi+\phi^{3}=0.
\end{equation}
In this paper a one-kink-like solution for Eq.(10) is constructed.
The direct Hirota method is applied for this purpose but some
modification of the method was realized.

Introduce a new dependent variable according to the Hirota method
\begin{equation} \label{EQ11}
\phi_{x}=\sigma \frac{F_{x}}{F},
\end{equation}
where $F=F(x,t)$ is a new unknown function, $\sigma$ is a constant
will be determined below. Now, Eq.(10) may be written in a form
$$
\frac{F_{xt}}{F}-2
\frac{F_{xt}F_{t}}{F^{2}}-\frac{F_{x}F_{tt}}{F^{2}}+ 2
\frac{F_{x}F_{t}^{2}}{F^{3}}-\frac{F_{xxx}}{F}+3
\frac{F_{x}F_{xx}}{F^{2}}-2 \frac{F_{x}^{3}}{F^{3}}+ 2 \beta
\frac{F_{xxt}}{F}- 2 \beta \frac{F_{xx}F_{t}}{F^{2}}-
$$
\begin{equation} \label{EQ12}
4 \beta \frac{F_{x}F_{xt}}{F^{2}}+ 4 \beta
\frac{F_{x}^{2}F_{t}}{F^{3}}+ \alpha \frac{F_{xt}}{F}- \alpha
\frac{F_{x}F_{t}}{F^{2}}- \sigma \frac{F_{x}}{F}+ \sigma^{2}
\frac{F_{x}^{3}}{F^{3}}=0.
\end{equation}

One looks for a partial solution, hence the next condition may be
accepted
\begin{equation} \label{EQ13}
(\sigma^{2}-2) \frac{F_{x}^{3}}{F^{3}}= 0.
\end{equation}
From this equation one obtains $\sigma=\sqrt{2}$. Usually, the
Hirota method may be used successfully if after condition Eq.(13)
the equation under study contains the second order term only. If
there are the third order terms, the equation may be solved by the
Hirota method for very limited and special cases \cite{13},
\cite{14}. In our problem after an accounting Eq.(13), two third
order terms are stayed in Eq.(12). It is clear, that an additional
condition is needed. Let write the such condition in a form
\begin{equation} \label{EQ14}
F_{t}+2\beta F_{x}=0.
\end{equation}
After an applying these two conditions to Eq.(12) it contains the
second order (bilinear) terms only:
$$
F_{xtt}F-2F_{xt}F_{t}-F_{x}F_{tt}-F_{xxx}F+3F_{x}F_{xx}+2 \beta
F_{xxt}F-2 \beta F_{xx}F_{t}-
$$
\begin{equation}\label{EQ15}
4 \alpha F_{x}F_{xt}+ \alpha F_{xt}F - \alpha F_{x}F_{t}-F_{x}F=0.
\end{equation}

From now all the steps in constructing of the one-kink-like solution
are standard ones for the Hirota method. As a result one obtains:
\begin{equation} \label{EQ16}
\phi(x,t)=\frac{1}{k}[1+ \tanh(\frac{kx- \omega t +
\eta_{0}}{\sqrt{2}})],
\end{equation}
where $k=- \frac{3}{4 \alpha \beta}$, $\omega = - \frac{3}{2 \alpha
}$. $\eta_{0}$ describes an initial position of solution. A
dispersion relation has a form
\begin{equation} \label{EQ17}
\omega^{2} - k^{2} - 2 \beta k \omega - \alpha \omega - 1= 0.
\end{equation}
As to condition (14) it takes a form
\begin{equation} \label{EQ18}
\omega = 2 \beta k.
\end{equation}
The last condition leads to relation $\alpha^{2} \beta^{2} >1$.

Substitute Eq.(16) into Eq.(10). Then for $\omega =- \frac{3}{2
\alpha}$ Eq.(16) will be a solution of Eq.(10) for
$k^{2}=\frac{1}{2}$ only. This condition is in agreement with
relation $\alpha^{2} \beta^{2} >1$. Namely, $\alpha^{2}
\beta^{2}=\frac{9}{8}$.

In this work a one-kink-like solution is constructed for the
Lorentz-violating $\phi^4$-theory equation of motion with
dissipation. It was done by some modification of the Hirota direct
method for nonlinear equation. This modification led to the special
conditions on coefficients of equation and parameters of solution.

\end{document}